\documentstyle[12pt,aasms4]{article}

 \mathcode`*="002A   
\def\lta{\mathbin{\lower 3pt\hbox
     {$\rlap{\raise 5pt\hbox{$\char'074$}}\mathchar"7218$}}}
\def\gta{\mathbin{\lower 3pt\hbox
     {$\rlap{\raise 5pt\hbox{$\char'076$}}\mathchar"7218$}}}

\begin{document}

\title{Comptonization and QPO Origins in Accreting Neutron Star Systems}

\author{Hyong C. Lee\altaffilmark{1} and Guy S. Miller\altaffilmark{2}}
\vspace*{0.3cm}
\affil{Department of Physics and Astronomy, Northwestern University,\\
       2145 Sheridan Road, Evanston, IL 60208, USA.}
\altaffiltext{1}{E-mail: hyongel@nwu.edu}
\altaffiltext{2}{E-mail: gsmiller@casbah.acns.nwu.edu}

\begin{abstract}

We develop a simple, time-dependent Comptonization model to probe the
origins of spectral variability in accreting neutron star systems.  In the
model, soft ``seed photons'' are injected into a corona of hot electrons,
where they are Compton upscattered before escaping as hard X-rays.  The
model describes how the hard X-ray spectrum varies when the properties of
either the soft photon source or the Comptonizing medium undergo small
oscillations.  Observations of the resulting spectral modulations can
determine whether the variability is due to (i)~oscillations in the
injection of seed photons, (ii)~oscillations in the coronal electron
density, or (iii)~oscillations in the coronal energy dissipation rate. 
Identifying the origin of spectral variability should help clarify how the
corona operates and its relation to the accretion disk.  It will also help
in finding the mechanisms underlying the various quasi-periodic
oscillations (QPO) observed in the X-ray outputs of many accreting neutron
star and black hole systems.  As a sample application of our model, we
analyze a kilohertz QPO observed in the atoll source 4U~1608-52.  We find
that the QPO is driven predominantly by an oscillation in the electron
density of the Comptonizing gas.

\end{abstract}

\keywords{accretion, accretion disks --- radiation mechanisms: thermal ---
stars: neutron --- X-rays: stars}

\section{Accretion Disk Coronae}

The X-ray spectra of accreting neutron stars and black holes in binary
systems typically appear to have at least two components  (for recent
reviews, see White, Nagase, \& Parmer~1995; and Tanaka \& Lewin~1995). 
The soft spectral component resembles blackbody radiation.  It is thought to
come from optically thick gas in the accretion disk and possibly, in the
case of an accreting neutron star, from gas near the stellar surface.  The
hard spectral component comes from hotter gas that fails to reach thermal
equilibrium with its radiation; the radiation intensity falls off with
increasing photon energy roughly as a power-law, $I_E\propto E^{-\alpha}$,
up to a cutoff photon energy $E_c$ beyond which the intensity drops
exponentially.  In the most rapidly accreting neutron star systems, the
Z-sources, the ``hard component'' cuts off at energies $\lta 10\,{\rm keV}$
and appears at least partially thermalized; the hard component extends to
higher energies in the atoll sources, which have lower accretion rates, and
in black hole candidate systems.  The multicomponent appearance of their
X-ray spectra suggests that the accretion flows around these objects may
be divided into a minimum of two substructures, one consisting of cooler,
dense gas that for convenience we shall call the ``disk'' component, and a
coronal component of hotter, tenuous gas.  Observations of rapid
variability in X-ray binary systems (see van der Klis 1995 for a review)
also support this picture of discrete physical substructures in the
accretion flow.  Fluctuations in the hard spectral component tend to be
much more pronounced than those in the soft component, and to be more
weakly correlated with fluctuations in the soft component on short
timescales than on longer timescales.

Comptonization has long been recognised as the probable source of the hard
spectral component in cosmic X-ray sources (Thorne \& Price 1975; Shapiro,
Lightman, \& Eardley 1976), since it naturally produces high-energy spectra
of approximately the right shape, and is the dominant interaction between
photons and electrons in hot, tenuous gases.  (Here we do not consider
neutron stars with surface magnetic fields $B_s\gta10^{12}\,{\rm Gauss}$,
as their extremely strong magnetic fields make the physics of their X-ray
spectra unlike that of other accreting neutron stars.)  The hot gas cools
by Compton scattering low-energy ``seed photons'' to higher energies, and
the Compton upscattered photons escape the system as the hard spectral
component.  The source of the seed photons is unclear at present, and is
likely to be different in black hole and neutron star systems.  In some
models for spectral formation in black hole systems (see, for example,
Haardt \& Maraschi 1993), soft photons from the optically thick disk seed
the Comptonization, but in others (e.g., Narayan, Yi, \& Mahadevan 1995)
the soft photons are due to cyclotron/synchrotron emission from the hot
corona itself.  In neutron star systems, cyclotron emission from the
neutron star magnetosphere can provide the seed photons, and it is
probable that Compton emission from hot magnetospheric gas dominates the
hard spectrum (Psaltis, Lamb, \& Miller~1995); nevertheless, the disk,
magnetospheric boundary layer, and stellar surface also may play important
roles.  The question of where seed photons come from depends on the
geometrical relation between the hot and cold flow substructures, which is
poorly known at best.  For example, in the case of accretion by a black
hole, advection-dominated disk models (see Narayan 1996 and references
therein) and the classic Shapiro, Lightman, and Eardley scenario have a
cold outer disk which becomes hot and optically thin to absorption within
some critical radius, so that the corona occupies the most central regions
of the flow, while the cold disk lies beyond it; however, models in which
the coronal gas sandwiches the colder disk have also received considerable
theoretical development (Liang \& Price~1977; Bisnovatyi-Kogan \&
Blinnikov~1977; and Galeev, Rosner, \& Vaiana~1979; more recent studies
emphasizing spectral formation include Haardt \& Maraschi~1993; Haardt,
Maraschi, \& Ghisellini~1997; and Dove, Wilms, \& Begelman~1997).  Given
the uncertainties, even very general information about the coronal
structure and soft photon source would be welcome, in black hole candidate
and in neutron star systems.

Rapid X-ray variability is now thought to hold vital clues to the nature of
the accretion flow ``engines'' in these systems.  In particular,
quasi-periodic oscillations have attracted attention from both observers
and theorists, since they are believed to contain otherwise inaccessible
information about the structure and operation of the central accretion flow.
(A quasi-periodic oscillation, or QPO, is enhanced variability in a
relatively narrow range of frequencies around a so-called ``centroid'' or
QPO frequency, which thus resembles but is not identical to periodic
modulation in the emission from a source.)  The QPO so far discovered in
X-ray binaries span a wide range of frequencies, from $\nu_{\rm
QPO}\lta1\,{\rm Hz}$ to over a kilohertz, and appear to belong to a number
of phenomenologically distinct categories, each of which presumably has a
different mechanism.  Proposed mechanisms for QPO tend to be situated
within or near the radiation-generating central regions of their accretion
flows, and include modulations of the flow at the inner edge of the disk or
at the magnetospheric boundary (for the classic ``magnetospheric beat
frequency'' models, see Alpar \& Shaham~1985, and Lamb et al.~1985; also
see Miller, Lamb, \& Psaltis~1997 for a beat frequency model applied to
the recently discovered kilohertz QPO), wave-like oscillations in the inner
disk (e.g., Chen \& Taam~1995, and references therein), and oscillations in
a coronal flow component (Fortner, Lamb, \& Miller 1989; Miller \& Lamb
1992).  Although leading candidate models have emerged for one or two of
the QPO categories, satisfactory explanations exist for very little of the
rich QPO phenomenology that has come to light over the past decade.  QPO
remain promising, and enigmatic.

Without assuming a detailed hydrodynamic model for the QPO, several attempts
have been made to extract general information about the physical
characteristics of the accretion flow and the QPO mechanism from the
spectral structure of the observed QPO; i.e., how the amplitude of the QPO
varies from one band of photon energies to the next, and the degree to
which the oscillation in a given band lags behind those in other bands. 
Early EXOSAT-based studies of the 20--50~Hz QPO in Z-sources (Hasinger~1987,
van der Klis~1987) showed that X-ray photons at higher energies lag behind
those at lower energies by several milliseconds, an effect which
immediately suggested the progression of photons from lower to higher
energies in a Comptonizing corona.  The discovery stimulated several groups
to perform calculations of time-dependent Comptonization in which the soft
photon injection rate underwent small oscillations about its average value
(Wijers, van~Paradijs, \& Lewin~1987; Stollman et al.~1987).  In addition
to injection rate oscillations, Stollman et al.~(1987) examined the effects
of small oscillations in the coronal electron density.  They concluded that
either type of modulation, in the soft photon source or in the coronal
electron density, could account for the observed hard lag; the data
available at the time were insufficient to discriminate between the two
possibilities.


Observations of accreting neutron star and black hole binary systems with
the Rossi X-ray Timing Explorer (RXTE) satellite have revealed unprecedented
detail in their X-ray variations, and motivate us to return to the issue of
how much can be learned about X-ray coronae and the origins of X-ray
variability through time-dependent Comptonization models.  In this paper we
develop a method of calculating spectral variations based on the
time-dependent Kompaneets equation that is far more computationally
efficient than the Monte Carlo methods employed by Wijers et al.~and
Stollman et al.  Our calculations permit us to introduce small oscillations
into {\em any\/} of the underlying properties of the Comptonization process
and study how they drive the emerging radiation.  Since the Kompaneets
approach is best for systems with scattering optical depths $\tau$ well
above unity, our method is more appropriate for accreting neutron star
systems (in which coronal optical depths are expected to be $\gta 5$) than
for black hole candidate systems (in which it is estimated $\tau\lta 1$). 
Using values for coronal parameters expected to be characteristic of
neutron star systems, we explore how X-ray spectral variations depend on
coronal properties and on the nature of the underlying driving modulation. 
We show that considerable differences exist between the spectral variations
produced by modulations in the soft photon injection, in the coronal
electron density, and in the coronal electron temperature.  These
differences should make it possible to fit observed spectral variations as
a superposition of the different types of modulation, and to obtain
constraints on QPO mechanisms and on system parameters such as the size of
the Comptonizing corona that are more reliable than those obtained by
considering only one type of modulation in isolation.

\section{Time-dependent Comptonization Model}

%
%

Since the properties of an accretion disk corona and its soft photon source
are so poorly known, we adopt the simplest possible description for them
consistent with our goal of physically interpreting variability in hard
X-ray spectra.  We ignore the effects of gradients in the electron density
and temperature of the corona, and treat the Comptonization process as if it
occurred in a completely homogeneous medium with electron density $n_e$ and
temperature $T$.  We do, however, allow these quantities to undergo small
oscillations as photons in the corona are Comptonized.  We further enforce
the assumption that the corona is homogeneous by stipulating that once
injected into the corona, any photon may escape from it with a probability
per unit time
\begin{equation}\label{escaperate}
\Gamma_{\rm esc}=c\sigma_Tn_e/N_{\rm esc} \equiv 1/(N_{\rm esc}t_c)\,,
\end{equation}
where $c$ is the speed of light and $\sigma_T$ is the Thomson cross section.
The escape rate $\Gamma_{\rm esc}$ depends only on the mean time between
collisions, $t_c$, and on the average number $N_{\rm esc}$ of collisions
suffered by a photon before its escape.  Thus the model omits any
information about the location of a photon within the corona, and all of
the properties of the corona are subsumed in the three quantities,
$T$, $t_c$, and $N_{\rm esc}$.

Our model also treats the injected photons simply.  Photons enter the
Comptonizing environment at a rate $\dot N$.  At the instant of its entry,
the probability that a photon has an energy between $E$ and $E+\delta E$ is
given by $f_{\rm inj}(E)\delta E$.  In this paper, we will take the
injected photon distribution
%
%
to have a blackbody spectrum:
$f_{\rm inj}(E) =AT_{\rm inj}^{-3}E^2(\exp[E/T_{\rm inj}]-1)^{-1}$, where
$A$ is a normalisation constant. (We write the temperature in units of
energy, so that Boltzmann's constant $k_B\equiv 1$.)  To study how
fluctuations in the injection process influence the spectrum of photons
escaping from the corona, we introduce small oscillations in $\dot N$ about
its average value.  We do not permit $f_{\rm inj}$ to vary.  Thus, we allow
for variability in the overall rate of photon injection, but ignore more
subtle modulations in the injected spectrum.

For nonrelativistic coronal temperatures ($T\ll m_ec^2$) and sufficiently
low photon energies ($E\ll m_ec^2$), the Kompaneets equation (Kompaneets
1957; see Psaltis and Lamb 1997 for a complete derivation of the equation
and a discussion of its limitations),
\begin{equation}\label{fullkompaneets}
t_{\rm c} \partial_{t} f = {1\over m_ec^2}\partial_{E}
\left[
-4TEf + E^2f + T\partial_E\left(E^2f\right)
\right]
+ t_c\dot{N}f_{\rm inj} -
{f\over N_{\rm esc}}\,,
\end{equation} 
describes the temporal development of the distribution $f$ of photons in the
Comptonizing medium.  The Kompaneets equation is essentially a photon
continuity equation in energy.  Its natural boundary conditions are that
the term in square brackets (the photon current) vanish as $E\rightarrow 0$
and as $E\rightarrow\infty$.

The spectrum of photons escaping from the Comptonizing medium is given by
\begin{equation}\label{foutdef}
f_{\rm out}(E)\equiv \Gamma_{\rm esc}f\,,
\end{equation}
so that modulations in the observed spectrum can be driven both by changes
in the spectrum $f$ of photons within the cloud and by changes in the rate
$\Gamma_{\rm esc}$ at which photons escape from the cloud.

To find how variations in the Comptonization process affect the escaping
spectrum, we assume that a given parameter in our model executes small
oscillations around its average value, and solve for the resulting
behavior of the escaping spectrum to first order in the oscillating
perturbation.  For example, to see how $f_{\rm out}$ reacts to variations in
the photon injection rate $\dot N$, we would write the injection rate as
the sum of a constant value $\dot N_0$ and a small oscillatory part:
\begin{equation}\label{inpertdef}
\dot N = \dot N_0 + \dot N_1 e^{-i\omega t}\,.
\end{equation}
Then to first order in the perturbation, the escaping photon spectrum is
given by
\begin{equation}\label{foutpertdef}
f_{\rm out}(E)\approx f_{\rm out\,0} + e^{-i\omega t}f_{\rm out\,1}\,;
\end{equation}
where $f_{\rm out\,0}$ is the escaping spectrum in the limit of a vanishing
perturbation,
\begin{equation}\label{foutinpert}
f_{\rm out\,1}
=
f_1/\left[t_c N_{\rm esc}\right]_0\,,
\end{equation}
and the perturbation amplitude $f_1$ of the coronal spectrum is given by the
linearised Kompaneets equation,
\begin{equation} \label{injection}
\left( \frac{1}{N_{\rm esc}} - i\omega t_c \right) f_1
+{1\over m_ec^2}\,{d\,\over dE}
\left[
4TEf_1 - E^2f_1 - T{d\,\over dE} E^2f_1
\right]
=
t_c f_{\rm inj}\dot{N_1}\,.
\end{equation}

In addition to perturbations in the soft photon input, we also consider two
types of perturbation in the corona itself.  Perturbations in the rate of
energy dissipation in the corona drive variations in its temperature:
\begin{equation}\label{tpertdef}
T = T_0 + T_1 e^{-i\omega t}\,.
\end{equation}
The escaping photon spectrum is still given by eq.~(\ref{foutinpert}), but
$f_1$ is now determined by
\begin{equation}\label{temperature}
\left( 
\frac{1}{N_{\rm esc}} - i\omega t_{\rm c} 
\right)
f_1
=
{1\over m_ec^2}\,{d\,\over dE}
\left[
-4E \left( T_1f_0 + T_0f_1 \right) 
+ E^2f_1
+ {d\,\over dE} E^2 \left( T_1f_0 + T_0f_1 \right)
\right]\,.
\end{equation} 

Perturbations in the supply of matter to the corona can cause its size $l$
or electron density $n_e$ to change, altering $t_c$ and $N_{\rm esc}$
We assume that $N_{\rm esc}$ is connected to the electron density $n_e$
through $N_{\rm esc}=\tau(\tau+1)$, where the scattering optical depth of
the corona is defined to be
$\tau\equiv\sigma_T n_e l$.  Thus,
\begin{equation}\label{nesctcrelation}
t_c={l\over c\tau}=
{2l\over c}\,\left(\left[4N_{\rm esc}+1\right]^{1/2}-1\right)^{-1}
\,.
\end{equation}
(This choice is motivated by simplicity; the exact form of the relationship
between $N_{\rm esc}$ and $t_c$ or $n_e$ depends on geometrical
assumptions, and on how the length scale $l$ is defined.  As an example,
for escape from the center of a homogeneous Comptonizing sphere, one finds
$N_{\rm esc}=3\pi^{-2}\tau[\tau+(4/3)]$ when
$l$ is defined as the radius $R$ of the sphere, but if one instead defines
$l\equiv3^{1/2}\pi^{-1}R \approx 0.6R$, the result is a relationship between
$N_{\rm esc}$ and $t_c$ close to the one we assume.)  Now the variation in
the escaping photon spectrum is given by
\begin{equation}\label{foutmatpert}
f_{\rm out\,1}
=
-f_{\rm out\,0}
\left(
{N_{\rm esc\,1}\over N_{\rm esc\,0}}
+{t_{c\,1}\over t_{c\,0}}
\right)
+ f_1/\left[t_c N_{\rm esc}\right]_0\,,
\end{equation}
where
\begin{equation} \label{matter}
\left( \frac{1}{N_{\rm esc\,0}} - i\omega t_{c\,0} \right) f_1
+
{1\over m_ec^2}\,{d\,\over dE}
\left[
4TEf_1 - E^2f_1 - T{d\,\over dE} E^2f_1
\right]
=
\dot N f_{\rm inj}t_{c\,1}
+
f_0{N_{\rm esc\,1}\over N_{\rm esc\,0}^2}
\end{equation}
governs the perturbation in the coronal spectrum.  Although in general we
could allow the parameters $t_c$ and $N_{\rm esc}$ to vary independently,
in this paper we will assume that the length scale $l$ of the corona
remains fixed (we choose $l=10^6\,{\rm cm}$ as characteristic of
circumstellar regions in neutron star systems), while the electron density
undergoes small oscillations.

This simple model for dynamic Comptonization allows us to study the
different spectral variabilities that arise from modulations in the supply
of coolant (the soft photons), the supply of heat, and the supply of matter
to the corona.  In the following section we will examine the characteristics
of each type of variability, paying special attention to traits that might
serve to distinguish one from another observationally.

\section{X-ray Variability: Numerical Results}


Each type of perturbation --- in photon injection, in coronal heating, and
in the electron density of the corona --- produces a different
characteristic variation in the spectrum of the escaping radiation.  By
identifying the superposition of perturbations that best matches the
observed spectral variations of a given source, we can learn more about
the physics of X-ray coronae, and in particular about their sources of
material and energy.  In this section we solve numerically for the
observable spectral variations produced by each of the basic types of
perturbation under a range of conditions.  We find that, as functions of
photon energy, the rms relative amplitude
\begin{equation}\label{relmagdef}
\xi(E)\equiv {1\over\sqrt 2}\,{\left|f_{\rm out\,1}\right|\over f_{\rm
out\,0}}
\end{equation}
and the phase
\begin{equation}\label{phasedef}
\theta(E)\equiv
\arctan
\left(
{{\rm Im}\,f_{\rm out\,1}\over{\rm Re}\,f_{\rm out\,1}}
\right)
\end{equation}
of each type of spectral variation differ considerably from the other types,
and so should be readily distinguishable.

We illustrate the main differences between the three types of spectral
variability with a typical model corona.  Coronal electrons with an average
temperature of $T_0=10\,{\rm keV}$ Comptonize 0.1~keV seed photons.  The
photons suffer on average $N_{\rm esc\,0}=100$ collisions before escaping
from the corona.  All the perturbations have frequencies
$\nu=\omega/(2\pi)=40\,{\rm Hz}$ and relative amplitudes of $10^{-2}$;
i.e., $\dot N_1/\dot N_0=10^{-2}$, $T_1/T_0=10^{-2}$, and $N_{\rm
esc\,1}/N_{\rm esc\,0}=10^{-2}$.

Figure~1 shows the relative amplitudes $\xi(E)$ (bold lines) and phases
$\theta(E)$ (lighter lines) produced by each perturbation.  Oscillations in
the seed photon injection rate $\dot N$ cause spectral variations with an
approximately energy-independent relative amplitude (bold short dashes);
since the residence time of photons in the corona, $t_{\rm
esc}=1/\Gamma_{\rm esc}$, is short compared to the oscillation period, the
spectral response is quasi-static.  To a good approximation, the emergent
spectrum simply rises and falls with the photon injection rate.  The phase
lag (light short dashes) is small and has a minimum near the seed injection
energy $T_{\rm inj}$; it rises monotonically with energy above $T_{\rm
inj}$.  The phase lag corresponds roughly to the time delay $\sim
t_c(m_ec^2/T)$ required by photons to upscatter from low energies $\sim
T_{\rm inj}$ to higher energies $\sim T_0$ (as may be estimated from the
Kompaneets equation~[\ref{fullkompaneets}]; also see Pozdnyakov, Sobol,
\&~Sunyaev~1983).  Temperature-driven spectral variations (solid lines) are
very different.  They have a minimum at photon energies near $3T_0$ and rise
sharply thereafter.  The associated phase variations also contrast strongly
with those due to oscillations in the photon injection rate, and reflect
the redistribution of photons from energies below $\sim 3T_0$ to higher
energies that occurs when the coronal temperature increases.  Oscillations
in the corona's optical depth (long dashes) alternately overpopulate and
depopulate the photon spectrum above $T_0$, resulting in a relative
amplitude that rises with photon energy near and above $T_0$, but less
steeply than one driven by coronal temperature oscillations.  The
spectrum below $\sim T_0$ falls as the optical depth rises.  Oscillations
at energies above $T_0$ are nearly out of phase with those below, except
for a phase lag that is caused by the upscattering time delay.  Thus, each
type of oscillation has its own distinctive signature.

In the following subsections, we systematically examine how each of the
spectral oscillation signatures depends on the frequency of the oscillation,
and on the average properties of the Comptonizing medium and soft photon
injection process.  The perturbed spectra depend most strongly on the
mean coronal electron temperature $T_0$, and on the ratio of the photon
residence time to the oscillation period, $\nu t_{\rm esc}$.   We choose
$T_0$ to be 5, 10, or 15~keV, characteristic of coronae in accreting
neutron star systems.  Injection energies $T_{\rm inj}$ are drawn from the
set \{0.1~keV, 0.5 keV, 1~keV\}.  Oscillation frequencies of
$\nu=40$, 100, and 1000~Hz are used to generate the perturbed spectra. 
Escaping photons experience an average of $N_{\rm esc\,0}=10$, 100, and 400
collisions in our calculations, so that the combination $\nu t_{\rm esc}$
ranges from $\sim10^{-3}$--$10^0$, and the spectral response goes from the
quasi-static limit to fully dynamic behavior.

\subsection{Perturbations in the Photon Injection Rate}

Suppose that the soft photon source oscillates slightly, driving small
oscillations in the emergent spectrum.  If the oscillation frequency
$\nu$ is small, $\nu t_{\rm esc}\ll 1$, the response of the spectrum is
close to the quasi-static ($\nu\rightarrow 0$) limit.  Thus, at all but the
highest frequencies, injection modulations simply cause the brightness of
the source to undergo small oscillations about its average value, with
little change in spectral shape.  The relative amplitude is therefore flat,
and the phase shifts are small (see Figs.~2a and~2b).  Spectral
oscillations at higher $\nu$ have larger phase lags and are more strongly
attenuated at energies away from $T_{\rm inj}$.
%
%

The dependence of the perturbed spectrum on the coronal temperature $T_0$
is illustrated in Fig.~2c.  The relative amplitude varies little with
changes in $T_0$.  Because the diffusion of photons in energy progresses
more rapidly in a hotter corona, the phase lags of the spectral variations
are smaller for a hotter corona than for a cooler one.  Figure~2d shows
that the amplitude of the spectral oscillations above 1~keV has almost no
dependence on the injection temperature.  The phase lags are somewhat more
sensitive to $T_{\rm inj}$, although the dependence is still not strong
(approximately logarithmic in $T_0/T_{\rm inj}$).  The phase delay is
minimised at energies near $T_{\rm inj}$, since photons escaping at
energies near $T_{\rm inj}$ typically have spent less time in the corona
than photons at more distant energies.

\subsection{Perturbations in Electron Temperature}

Oscillations in the coronal electron temperature force photons from lower
to higher energies.  The spectral oscillations are thus anticorrelated with
the temperature oscillations at low energies, and correlated at higher
energies (Fig.~3a).  The energy at which the behavior changes from
anticorrelation to correlation corresponds to a minimum in the relative
amplitude.

When the Comptonized spectrum is very nonthermal (i.e., when the
``y-parameter'' $y\equiv 4N_{\rm esc\,0}T_0/(mc^2)\ll 1$; see Pozdnyakov et
al.~1983), the minimum occurs at energies somewhat above $T_{\rm inj}$, but
considerably lower than $T_0$.  Fig.~3b shows that the minimum moves toward
an asymptotic position at $3T_0$ in the case of more thermalized spectra,
for which the Wien peak at $2T_0$ is more prominent (these correspond to
$y\gg 1$).  This follows from the fact that in the neighborhood of the Wien
peak, we expect the time-averaged spectrum to be $f_0\approx c_{\rm
norm}E^2T_0^{-3} e^{-E/T_0}$, where $c_{\rm norm}$ is a normalisation
constant independent of $T_0$.  The $T_0$-dependence follows from the shape
of the thermal peak, and the need to maintain a constant photon number
(constant normalisation) at different electron temperatures.  In the
quasi-static limit the relative spectral variation is then $T_1e^{-i\omega
t}d\ln f_0/dT_0$, and so the relative amplitude is
$\xi\approx|(-3/T_0+E/T_0^2)T_1|$.  The approach to the quasi-static,
thermalized limit is seen in Fig.~3c.  The minimum in the relative
amplitude occurs at $6\,{\rm keV}\approx 1.2T_0$ for the relatively
unthermalized case $T_0=5\,{\rm keV}$, and has reached $33\,{\rm
keV}\approx 2.2T_0$ in the more thermalized case $T_0=15\,{\rm keV}$.

As Fig.~3d shows, the influence of the photon injection temperature on the
spectral oscillations is minor.  It is slightly easier for photons
introduced at higher energies to form a Wien peak, all other things being
equal.  Consequently, cases with higher $T_{\rm inj}$ are a little better
thermalized than cases with lower $T_{\rm inj}$, and have amplitude minima
at slightly higher energies.

\subsection{Perturbations in Electron Density}

When the coronal electron density oscillates, the spectral variations at
energies near $T_{\rm inj}$ are nearly anticorrelated with the driving
oscillations.  The anticorrelation occurs because an increase in optical
depth both suppresses the rate at which photons escape from the cloud and
allows photons to evolve to energies more distant from $T_{\rm inj}$.  The
latter effect is dominant at energies $E\gta T_0$, where the spectral
oscillation is nearly in phase with the electron density oscillation, except
for a small phase lag due mainly to the time required for photons to diffuse
out of the corona.

The dependence of the spectral variations on driving frequency is shown in
Fig.~4a.  Unsurprisingly, the relative amplitudes of the variations only
begin to betray their frequency dependence when the oscillations are
fastest, and the spectral response emerges from the quasi-static limit. 
Since the time lag between the responses in different energy channels has
little dependence on the frequency of the driving oscillations in the
quasi-static limit, the phase lag at a high photon energies grows
approximately linearly with frequency. As the mean electron density
increases, the time lag increases, along with the escape time $t_{\rm
esc}$.  This is seen in the phase lags of Fig.~4b.

Figure~4c shows how the spectral oscillations depend on the mean coronal
electron temperature.  As one would expect, the minimum in the relative
amplitude, which corresponds to the point where the spectral
oscillations go from their low-energy anticorrelation with the coronal
density oscillations to their high-energy correlation, increases with
coronal temperature.  When $T_{\rm inj}$ is increased, both the phase lag
and relative amplitude curves shift upward slightly in energy (Fig.~4d).

\section{Preliminary Application to Kilohertz QPO}

To illustrate how time-dependent Comptonization calculations can be
utilised in data interpretation, we present an informal analysis of an
850~Hz QPO in 4U~1608-52, based on published observations of the
energy-dependent QPO phase (Vaughan et al.~1997) and relative amplitude
(Berger et al.~1996).  With the assumption that oscillations in the soft
photon source drive the QPO, Vaughan et al.~estimated from the observed
phase lags that the size scale of the coronal gas is between a few
kilometers and a few tens of kilometers.

We model the observed spectral variability as a superposition of ``basis
variabilities'' driven by oscillations (i)~in the soft photon source,
(ii)~in the electron density of the Comptonizing corona, and (iii)~in the
temperature of the coronal electrons.  This approach allows the
possibility that the energy dependence of the observed oscillation phase
might derive from the underlying phase relationships of the basis
variabilities and their different energy dependences.  For example, a
photon injection oscillation might combine with small temperature
oscillation that lags it by a quarter cycle.  Since the size of the
temperature-induced spectral oscillation grows with increasing photon
energy, the resulting spectrum would have phase lags that increase steeply
with energy, but are unrelated to the time it takes for photons to
propagate from lower to higher energies.  To safely interpret the phase
lags, one should account consistently for the relative amplitude data and
the time-averaged spectrum.  It is also important that coronal emission
dominates the spectral range under consideration.  Contamination by
(steady) noncoronal emission will reduce the relative amplitude at low
photon energies below the model's predictions.

Since time-averaged spectra for 4U~1608-52 during the QPO observations are
not yet publicly available, the coronal parameters we derive must be taken
as preliminary, subject to future revision.  For example, we found that
acceptable fits to the QPO observations could be produced with a wide
range of coronal temperatures $T_0$.  Here, for purposes of illustration,
we assumed a temperature $T_0=10\,{\rm keV}$, and an injection spectrum
with $T_{\rm inj0}=0.1\,{\rm keV}$.  We adjusted the remaining parameters
by hand until an acceptable fit was obtained.  Our ``best fit'' appears
in Fig.~5.  The coronal parameters for the fit are: $l=6\,{\rm km}$ and
$N_{\rm esc 0}=20$ (or equivalently, $\tau=4$).  The composition of the
driving perturbation is $n_{e1}/n_{e0}=0.32$, $\dot N_1/\dot
N_0=-0.5n_{e1}/n_{e0}$, and $T_1/T_0=-(0.37+0.08i)n_{e1}/n_{e0}$.  Although
there is considerable latitude in acceptable admixtures of the injection
and temperature perturbations, we found that {\em to successfully match the
relative amplitude data, the primary source of the QPO has to be an
oscillation in the coronal electron density.  Oscillations dominated by
injection or coronal temperature oscillations produce unacceptable fits,
even at the qualitative level.}

\section{Conclusions}

Since the spectral variations driven by oscillations in the corona's supply
of soft photons, heat, and material each bear their own distinctive
characteristics, one can determine the nature of the oscillations driving
QPO.  Our sample application to a kilohertz QPO observed in 4U~1608-52
indicates that the main driver behind the QPO is an oscillation in coronal
electron density.  In agreement with earlier studies (Vaughan et al.~1997),
we find that the size of the corona is comparable to estimates of the
neutron star radius, supporting the notion that kilohertz QPO originate
very close to the accreting neutron star.

%
%

\acknowledgements

This work was supported in part by NASA grants NAGW-2935 and NAG5-3396.

\clearpage

\def\aa{A\&A}
\def\apj{ApJ}
\def\araa{ARA\&A}
\def\jetp{Sov.~Phys.-JETP}
\def\mnras{MNRAS}
\def\nat{Nature}
\def\pasj{PASJ}
\def\ssr{Space Sci.~Rev.}

\clearpage

\begin{figure}
\caption {
When the Comptonization process is disturbed in different ways, distinctive
patterns of spectral variability result.  Shown are the spectral variations
driven by 1\% oscillations in (i)~the soft photon injection rate (short
dashed curves),  (ii)~the coronal electron temperature (unbroken curves),
and (iii)~the coronal electron density (long dashed curves).  The
driving oscillations are all at 40~Hz.  Bold lines show the rms relative
amplitudes of the spectral oscillations as functions of photon energy.  An
rms relative amplitude of 1\% corresponds to a 1\% oscillation in photon count
rate about its average value.  Light lines show the phase lag between the
spectral oscillation and the physical oscillation driving it.  Each
spectral variation is produced in a ``standard'' corona with the following
time-averaged parameters: $N_{\rm esc 0}=100$, $T_{\rm inj}=0.1\,{\rm
keV}$, $T_{\rm e}=10\,{\rm keV}$, and $l_{\rm size}=10^6 {\rm cm}$.
}
\end{figure}

\begin{figure}
\caption
{
Spectral variations driven by 1\% oscillations in the photon injection
rate are illustrated in Figs.~2a--2d.  For each variation, a heavy
curve denotes the relative amplitude of the spectral oscillation, while
the phase lag is depicted by a light curve.  Each panel shows how the
spectral variations depend on a particular parameter (the oscillation
frequency, the average number of collisions suffered by an escaping
photon, the coronal temperature, or the temperature of the injected
photons), while the other parameters are kept at the values employed in
Fig.~1.  Fig.~2a compares the spectral responses at different driving
oscillation frequencies.  Variations induced by photon injection
oscillations with frequencies of 40~Hz (unbroken curves), 100~Hz (short
dashed curves), and 1~kHz (long dashed curves) are shown.  Fig.~2b
displays variations in the output spectra of coronae having $N_{\rm esc
0}=10$ (short dashed curves), 100 (unbroken curves), and 400 (long dashed
curves).  Fig.~2c shows variations in the output spectra of coronae with
electron temperatures of 5~keV (short dashed curves), 10~keV (unbroken
curves), and 15~keV (long dashed curves).  Fig.~2d contains spectral
variations with several different input photon temperatures: 0.1~keV
(unbroken curves), 0.5~keV (short dashed curves), and 1~keV (long dashed
curves).
}
\end{figure}

\begin{figure}
\caption
{
Spectral variations driven by 1\% oscillations in the coronal temperature.
As in Fig.~2, each panel shows how the spectral oscillations depend on the
underlying physical properties of the Comptonizing corona and its supply
of soft photons.
}
\end{figure}

\begin{figure}
\caption
{
Spectral variations driven by 1\% oscillations in the coronal electron
density.  Each panel shows how the spectral oscillations depend on the
underlying physical properties of the Comptonizing corona and its supply
of soft photons.  The plotting conventions are the same as in Fig.~2.
}
\end{figure}

\begin{figure}
\caption
{
Sample fit to an 850~Hz QPO observed in 4U~1608-52.  The data for the
relative amplitudes are from Berger et al.~1996, and the phase data are
from Vaughan et al.~1997.  The relative amplitude data imply that the QPO
is caused mainly by an oscillation in the coronal electron density, rather
than by oscillations in the soft photon source or in the coronal
temperature.  Details of the other coronal parameters are given in the
text.
}
\end{figure}

\clearpage

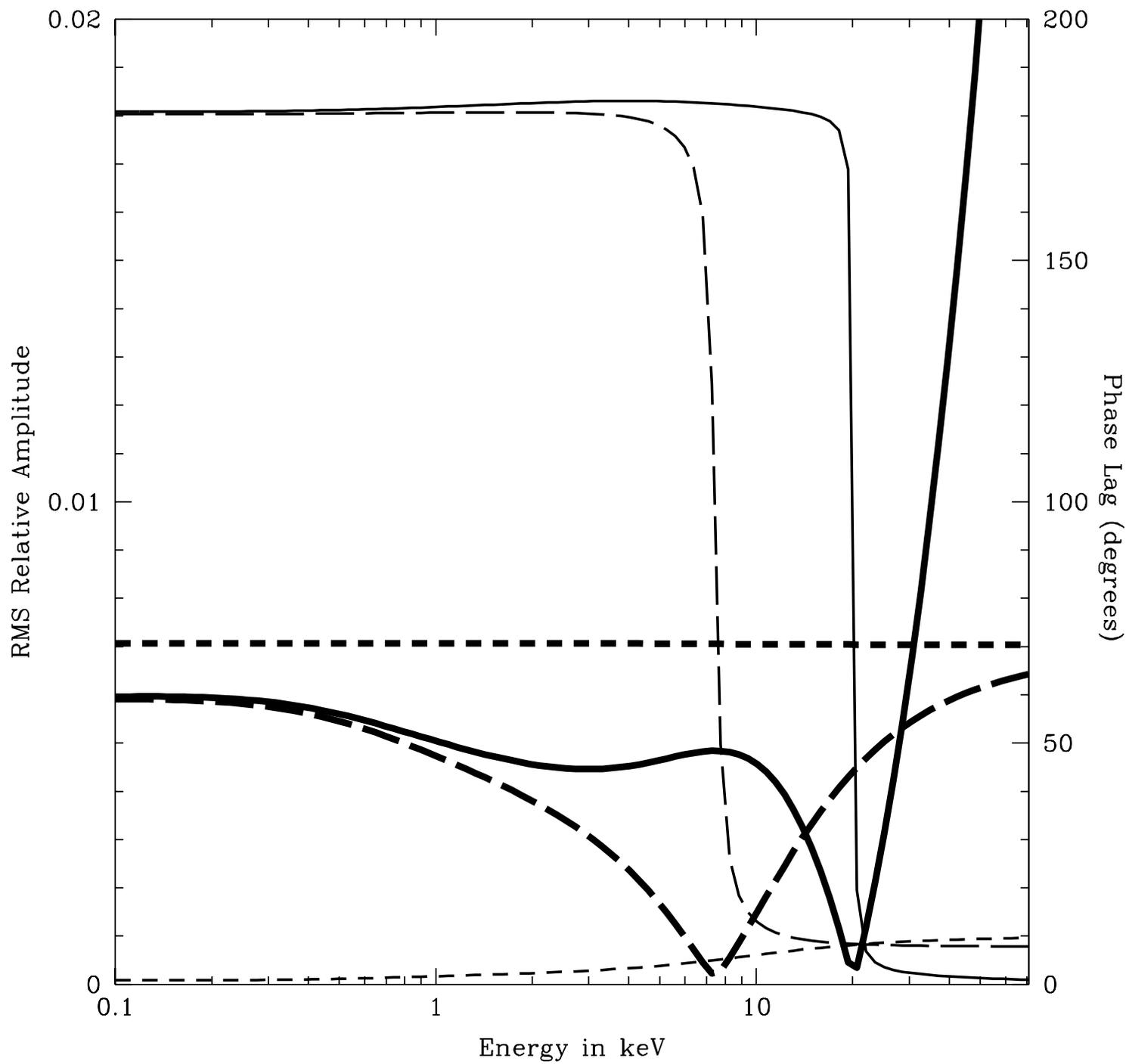
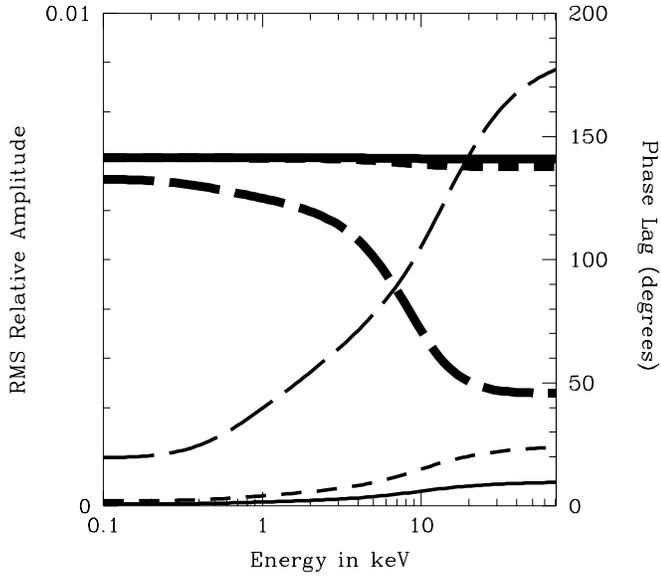
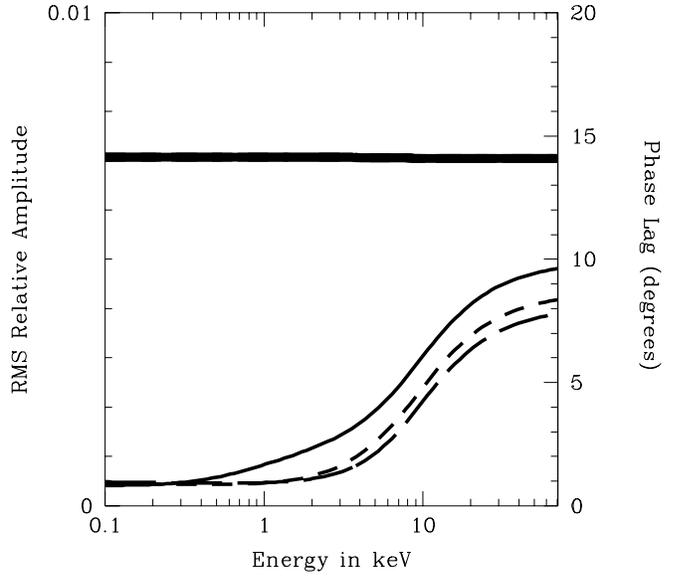
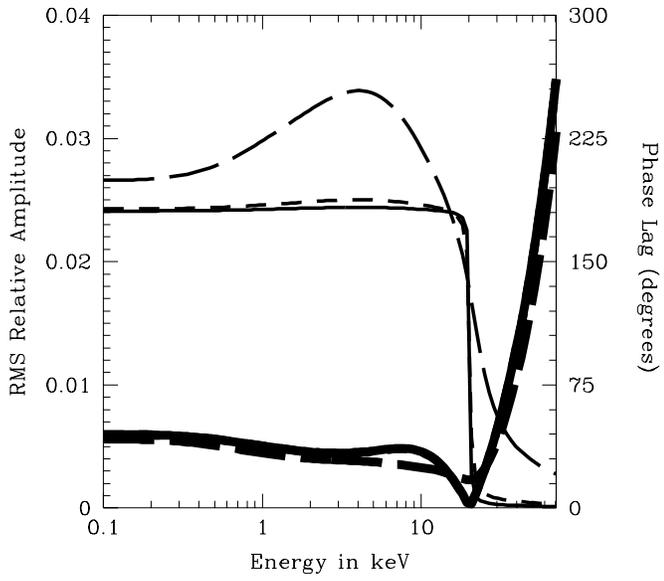
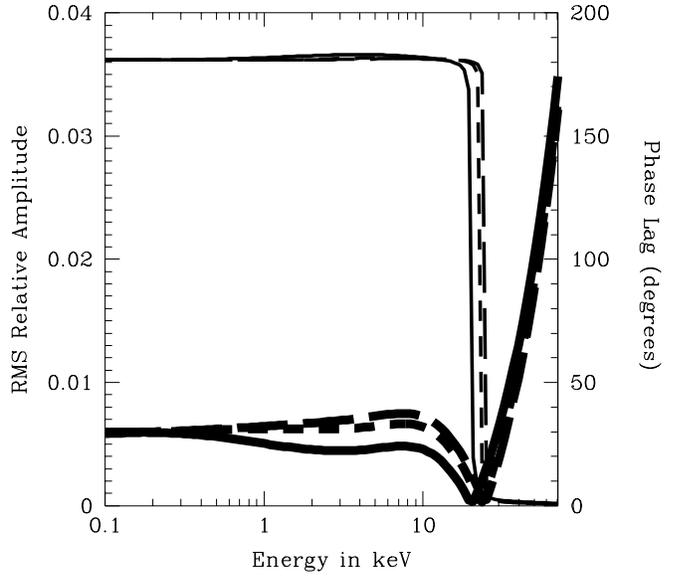
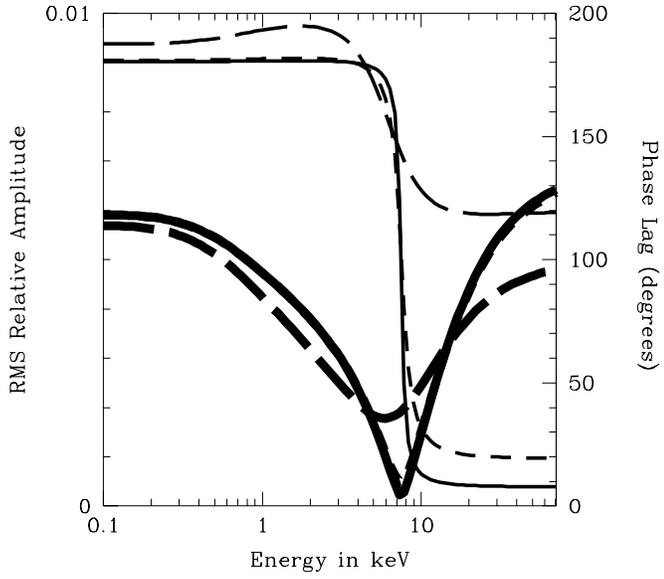
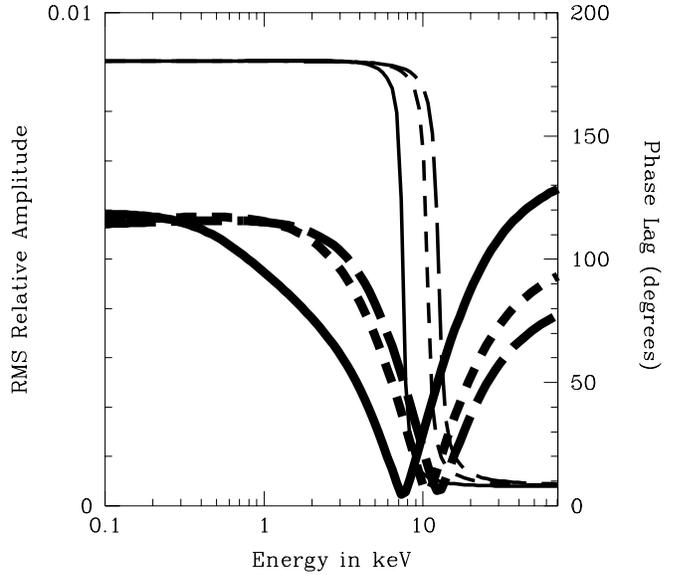
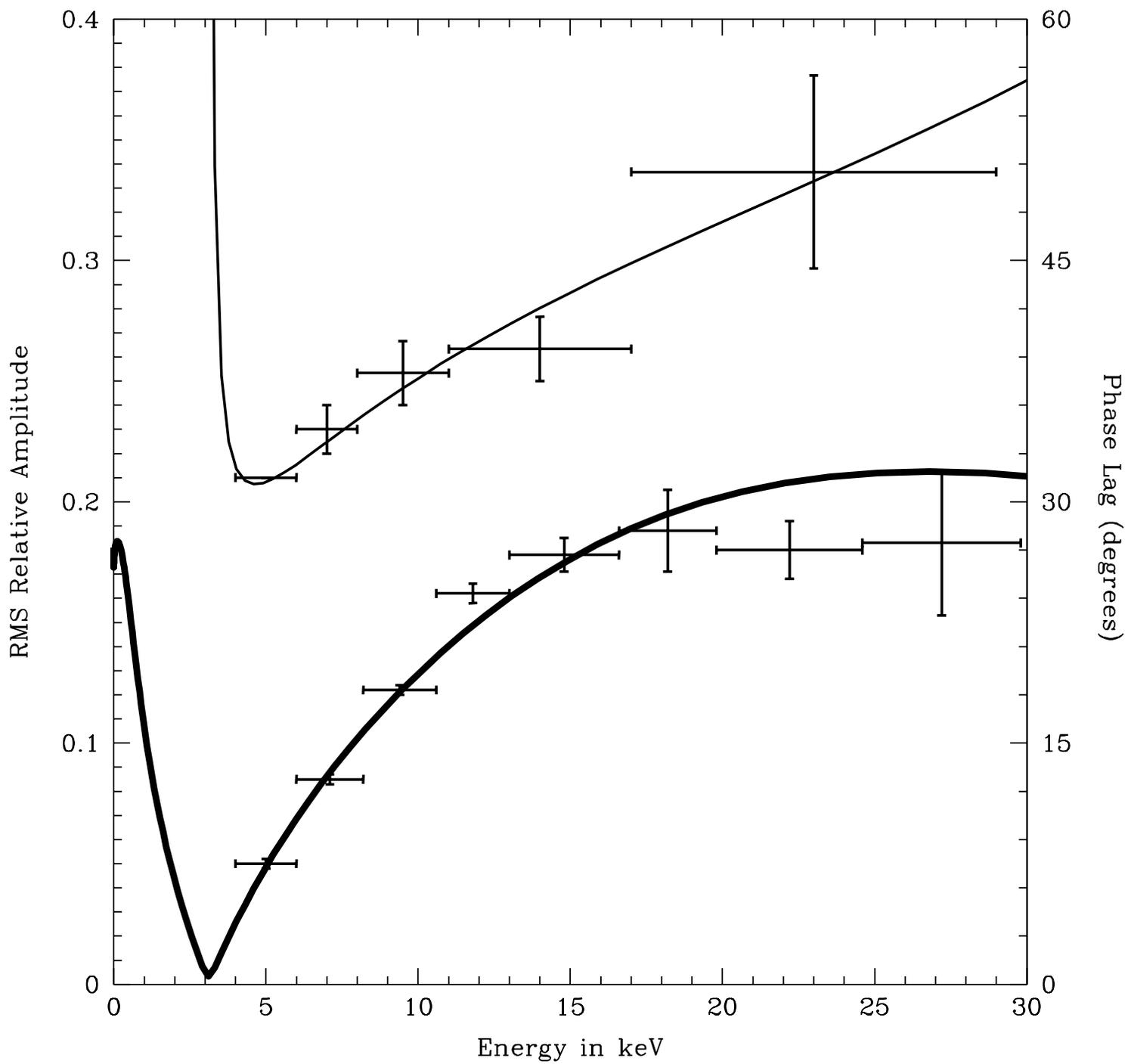
\end{document}